\documentclass [10pt]{article}
\usepackage[left=2cm,top=2.50cm,right=2cm,bottom=2.50cm]{geometry}
\usepackage{mathrsfs}
\usepackage{physics}
\newcommand*\rfrac[2]{{}^{#1}\!/_{#2}}

\usepackage{amsmath,amssymb,latexsym,color,cancel,graphicx,bbm,colortbl}
\usepackage[english]{babel}
\usepackage[latin1]{inputenc}
\usepackage{ragged2e}
\usepackage{cite}

\begin{document}
\date{}

\title{Exact solution of the isotropic and anisotropic Hamiltonian of two coupled harmonic oscillators}
\author{J. C. Vega$^{a}$, D. Ojeda-Guill\'en$^{b}$\footnote{{\it E-mail address:} dojedag@ipn.mx}, and R. D. Mota$^{c}$} \maketitle

\begin{minipage}{0.9\textwidth}
\small $^{a}$ Escuela Superior de F{\'i}sica y Matem\'aticas, Instituto Polit\'ecnico Nacional,
Ed. 9, U.P. Adolfo L\'opez Mateos, Alc. Gustavo A. Madero, C.P. 07738 Ciudad de M\'exico, Mexico.\\

\small $^{b}$ Escuela Superior de C\'omputo, Instituto Polit\'ecnico Nacional,
Av. Juan de Dios B\'atiz esq. Av. Miguel Oth\'on de Mendiz\'abal, Col. Lindavista,
Alc. Gustavo A. Madero, C.P. 07738, Ciudad de M\'exico, Mexico.\\

\small $^{c}$ Escuela Superior de Ingenier{\'i}a Mec\'anica y El\'ectrica, Unidad Culhuac\'an,
Instituto Polit\'ecnico Nacional, Av. Santa Ana No. 1000, Col. San
Francisco Culhuac\'an, Alc. Coyoac\'an, C.P. 04430, Ciudad de M\'exico, Mexico.\\
\end{minipage}

\section*{Abstract}
We study the isotropic and anisotropic Hamiltonian of two coupled harmonic oscillators from an algebraic approach of the $SU(1,1)$ and $SU(2)$ groups. In order to obtain the energy spectrum and eigenfunctions of this problem, we write its Hamiltonian in terms of the boson generators of the $SU(1,1)$ and $SU(2)$ groups. We use the one boson and two boson realizations of the $su(1,1)$ Lie algebra, and the one boson realization of the $su(2)$ Lie algebra to apply three tilting transformations to diagonalize the original Hamiltonian. These transformations let us to obtain the exact solutions of the isotropic and the anisotropic cases, from which the particular expected results are obtained for the cases where the coupling is neglected.

\section{Introduction}

Of all the problems that have contributed to our understanding of the universe around us, perhaps the simple harmonic oscillator is the most important. It has been a fundamental piece in the development of all branches of physics and its applications have extended to many different branches of science \cite{Bloch}. The mathematical simplicity of harmonic oscillators is contrasted by the enormous variety of methods and techniques that can be used to describe them. However, harmonic oscillators are not always disconnected from each other, and this is where the importance of studying coupled harmonic oscillators lies.

The study of coupled oscillators has found its application in nonlinear systems, as can be seen in Refs. \cite{Fano,Schweber,Estes,Fetter,Kim,Han,Iachello,Han2,Jakub}. One of the most studied characteristics of coupled oscillators is their quantum entanglement \cite{Joshi,Paz,Galve,Fillaux}, which has been applied to biophysics, quantum cryptography, quantum-coding, quantum computing algorithms, quantum state teleportation, among others \cite{Romero,Fuller,Halpin,Ekert,Bennett,Shor,Samuel}. In particular, in Ref. \cite{Makarov} the quantum entanglement of oscillators with position-position coupling is studied and the solution to the nonstationary Schr\"odinger equation is found.

The aim of this work is to study and exactly solve isotropic and anisotropic Hamiltonian of two coupled harmonic oscillators from an algebraic approach. We obtain its eigenfunctions and the energy spectrum by using similarity transformations in terms of the $su(1,1)$ and $su(2)$ Lie algebras.

This work is organized as it follows. In Sec. $2$ we write the Hamiltonian of the two isotropic coupled oscillators in terms of the bosonic operators $a$ and $b$. Then, we introduce Schwinger realizations in terms of one and two bosons of the $su(1,1)$ and $su(2)$ Lie algebras and apply three tilting transformations in terms of these realizations to the Hamiltonian to diagonalize it. These similarity transformations let us exactly solve the problem obtaining the eigenfunctions and the energy spectrum. In. Section $3$ we apply the three tilting transformations to the anisotropic case of our Hamiltonian in order to obtain the most general solution to our problem.
Finally, we give some concluding remarks.

\section{Hamiltonian of two coupled harmonic oscillators: The isotropic case}

A Hamiltonian which describes the behavior of a system of two coupled harmonic oscillators with identical free frequencies can be seen as a Hamiltonian which coupling is of the position-position type. It is defined as
\begin{equation}
    H_{xy} = \frac{1}{2m} \left( p_x^2 + p_y^2 \right) + \frac{m \omega^2}{2} \left( x^2 + y^2 \right) + 2\kappa m \omega xy,
\end{equation}
where $\kappa$ is a constant that describes the strength of the coupling. This Hamiltonian can be written in terms of the bosonic operators as
\begin{equation}
    H_{xy} = \hbar \omega \left( a^{\dag}a + \frac{1}{2} \right) + \hbar \omega \left( b^{\dag}b + \frac{1}{2} \right) + \hbar \kappa (a^{\dag} + a) (b^{\dag} + b).
\end{equation}

In order to study the system of two coupled harmonic oscillators with identical free frequencies we propose the following more general Hamiltonian (with $\hbar = 1$)
\begin{equation}
    H = \omega (a^{\dagger} a + b^{\dagger} b + 1) + \lambda e^{-i \psi} (a^{\dagger} b^{\dagger} + a^{\dagger} b) + \lambda e^{i \psi} (b^{\dagger} a + b a). \label{Hamiltonian}
\end{equation}
Using the two-bosons Schwinger realizations of the $SU(2)$ and $SU(1,1)$ Lie algebras of equations (\ref{su2}) and (\ref{su11ab}), we can write this Hamiltonian as
\begin{equation}
    H = 2 \omega K_0 + \lambda e^{-i \psi} (K_+ + J_+) + \lambda e^{i \psi} (K_- + J_-).
\end{equation}

From now on we will diagonalize the Hamiltonian $H$ by applying three consecutive tilting transformations, expressed in terms of the $SU(1,1)$ and $SU(2)$ displacement operators (see Appendix). Then, we first apply a tilting transformation to the stationary Schr\"odinger equation $H\Psi=E\Psi$ as it follows \cite{Gerryberry,Nos1,Nos2}
\begin{equation}
D^{\dagger}(\xi)HD(\xi)D^{\dagger}(\xi)\Psi=ED^{\dagger}(\xi)\Psi.
\end{equation}
Here, $D(\xi)$ is the $SU(1,1)$ displacement operator defined as $D(\xi) = \exp(\xi K_+ - \xi^{*} K_-)$, with $\xi = - \rfrac{1}{2} \tau e^{-i \phi_\xi}$. Thus,
if we define the tilted Hamiltonian $H'$ as $H'=D^{\dagger}(\xi)HD(\xi)$ and its wavefunctions $\Psi'$ as $\Psi'=D^{\dagger}(\xi)\Psi$, from the similarity transformations (\ref{st1}) and (\ref{st2}) of Appendix B we obtain
\begin{eqnarray}
        H' &=& 2\omega \left[ (2 \beta_\xi + 1) K_0 + \frac{\alpha_\xi \xi}{2 \abs{\xi}} K_+ + \frac{\alpha_\xi \xi^{*}}{2 \abs{\xi}} K_- \right] \nonumber\\
        &&+ \lambda e^{-i \psi} \left[ \left[ \frac{\xi^*}{\abs{\xi}} \alpha_\xi K_0 + \beta_\xi \left( K_+ + \frac{\xi^*}{\xi} K_- \right) + K_+ \right] + \left[ \frac{\xi^*}{\abs{\xi}} \alpha_\xi K_{-}^{(b)} + \frac{\xi}{\abs{\xi}} \alpha_\xi K_{+}^{(a)} + (2 \beta_\xi + 1) J_+ \right] \right] \nonumber\\
        &&+ \lambda e^{i \psi} \left[ \left[ \frac{\xi}{\abs{\xi}} \alpha_\xi K_0 + \beta_\xi \left( K_- + \frac{\xi}{\xi^*} K_+ \right) + K_- \right] + \left[ \frac{\xi^*}{\abs{\xi}} \alpha_\xi K_{-}^{(a)} + \frac{\xi}{\abs{\xi}} \alpha_\xi K_{+}^{(b)} + (2 \beta_\xi + 1) J_- \right] \right],
\end{eqnarray}
where $\alpha_\xi = \sinh{(2\abs{\xi})}$ and $\beta_\xi = \rfrac{1}{2} [ \cosh{(2\abs{\xi})} - 1]$. Notice that in this expression we have introduced $\{K_{\pm}^{a},K_{\pm}^{b}\}$, which are the one-boson ladder operators for the $SU(1,1)$ realization (see Eq. (\ref{su11a}) of Appendix A). This tilted Hamiltonian $H'$ can be written as
\begin{eqnarray}
        H' &=& \left[ 2 \omega (2 \beta_\xi + 1) + \frac{\xi^*}{\abs{\xi}} \alpha_\xi \lambda e^{-i \psi} + \frac{\xi}{\abs{\xi}} \alpha_\xi \lambda e^{i \psi} \right] K_0 \nonumber\\
        &&+ \left[ \frac{\omega \alpha_\xi \xi}{\abs{\xi}} + (\beta_\xi + 1) \lambda e^{-i \psi} + \frac{\beta_\xi \xi}{\xi^*} \lambda e^{i \psi} \right] K_+ + \left[ \frac{\omega \alpha_\xi \xi^*}{\abs{\xi}} + \frac{\beta_\xi \xi^*}{\xi} \lambda e^{-i \psi} + (\beta_\xi + 1) \lambda e^{i \psi} \right] K_- \nonumber\\
        &&+ \frac{\xi^*}{\abs{\xi}} \alpha_\xi \lambda (K_{-}^{(b)} e^{-i \psi} + K_{-}^{(a)} e^{i \psi}) + \frac{\xi}{\abs{\xi}} \alpha_\xi \lambda (K_{+}^{(a)} e^{-i \psi} + K_{+}^{(b)} e^{i \psi}) + (2 \beta_\xi + 1) \lambda (J_+ e^{-i \psi} + J_- e^{i \psi}).
\end{eqnarray}
If we choose the values of the coherent state parameters $\tau$ and $\phi_\xi$ such that the coefficients of the operators $K_{\pm}$ vanish, we find
\begin{equation}
    \tau = \tanh^{-1}{\left( \frac{\lambda}{\omega} \right)}, \hspace{0.3cm} \phi_\xi = \psi.
\end{equation}
This let us obtain the following reduced expression of the Hamiltonian $H'$
\begin{eqnarray}
        H' &=& 2[\omega \cosh{(\tau)} - \lambda \sinh{(\tau)}] K_0 + \lambda \cosh{(\tau)} (J_+ e^{-i \psi} + J_- e^{i \psi}) \nonumber\\
        &&- \lambda \sinh{(\tau)} (K_{+}^{(b)} + K_{-}^{(b)}) - \lambda \sinh{(\tau)} (K_{+}^{(a)} e^{-2i \psi} + K_{-}^{(a)} e^{2i \psi}).
\end{eqnarray}
Using the expressions $\cosh{(\tau)} = \frac{\omega}{\sqrt{\omega^2 - \lambda^2}}$ and $\sinh{(\tau)} = \frac{\lambda}{\sqrt{\omega^2 - \lambda^2}}$ we arrive to
\begin{eqnarray}
        H' &=& 2 \sqrt{\omega^2 - \lambda^2} K_0 + \frac{\omega \lambda}{\sqrt{\omega^2 - \lambda^2}} (J_+ e^{-i \psi} + J_- e^{i \psi}) \nonumber\\
        &&- \frac{\lambda^2}{\sqrt{\omega^2 - \lambda^2}} (K_{+}^{(b)} + K_{-}^{(b)}) - \frac{\lambda^2}{\sqrt{\omega^2 - \lambda^2}} (K_{+}^{(a)} e^{-2i \psi} + K_{-}^{(a)} e^{2i \psi}).
\end{eqnarray}
Now, following a similar procedure, we can apply to $H'$ a second tilting transformation. Thus, in terms of the $SU(2)$ displacement operator $D(\chi) = \exp(\chi J_+ - \chi^{*} J_-)$, we can define $H''$ and $\Psi''$ as $H''=D^{\dagger}(\chi)H'D(\chi)$ and $\Psi''=D^{\dagger}(\chi)\Psi'$, respectively. Hence, the similarity transformations of Eqs. (\ref{st3})-(\ref{st5}) of Appendix B let us to obtain
\begin{equation}
    \begin{gathered}
        H'' = 2 \sqrt{\omega^2 - \lambda^2} K_0 \\
        + \frac{\omega \lambda}{\sqrt{\omega^2 - \lambda^2}} \left[ e^{-i \psi} \left[ - \frac{\chi^*}{\abs{\chi}} \alpha_\chi J_0 + \beta_\chi \left( J_+ + \frac{\chi^*}{\chi} J_- \right) + J_+ \right] + e^{i \psi} \left[ - \frac{\chi}{\abs{\chi}} \alpha_\chi J_0 + \beta_\chi \left( J_- + \frac{\chi}{\chi^*} J_+ \right) + J_- \right] \right] \\
        - \frac{\lambda^2}{\sqrt{\omega^2 - \lambda^2}} \left[ \left[ \frac{K_{-}^{(b)}}{2} \left( \cos{(2 \abs{\chi})} + 1 \right) - \frac{\chi^*}{2 \abs{\chi}} \sin{(2 \abs{\chi})} K_- - \frac{\chi^*}{2 \chi} \left( \cos{(2 \abs{\chi})} - 1 \right) K_{-}^{(a)} \right] \right. \\
        \left. + \left[ \frac{K_{+}^{(b)}}{2} \left( \cos{(2 \abs{\chi})} + 1 \right) - \frac{\chi}{2 \abs{\chi}} \sin{(2 \abs{\chi})} K_+ - \frac{\chi}{2 \chi^*} \left( \cos{(2 \abs{\chi})} - 1 \right) K_{+}^{(a)}  \right] \right] \\
        - \frac{\lambda^2}{\sqrt{\omega^2 - \lambda^2}} \left[ e^{-2i \psi} \left[ \frac{K_{+}^{(a)}}{2} \left( \cos{(2 \abs{\chi})} + 1 \right) + \frac{\chi^*}{2 \abs{\chi}} \sin{(2 \abs{\chi})} K_+ - \frac{\chi^*}{2 \chi} \left( \cos{(2 \abs{\chi})} - 1 \right) K_{+}^{(b)} \right] \right. \\
        \left. + e^{2i \psi} \left[ \frac{K_{-}^{(a)}}{2} \left( \cos{(2 \abs{\chi})} + 1 \right) + \frac{\chi}{2 \abs{\chi}} \sin{(2 \abs{\chi})} K_- - \frac{\chi}{2 \chi^*} \left( \cos{(2 \abs{\chi})} - 1 \right) K_{-}^{(b)} \right] \right].
    \end{gathered}
\end{equation}
where $\alpha_\chi = \sin{(2\abs{\chi})}$ and $\beta_\chi = \rfrac{1}{2} [ \cos{(2\abs{\chi})} - 1]$. \\
We can remove the dependence of the Hamiltonian $H''$ on the operators $J_{\pm}$ if the coherent state parameters are chosen as
\begin{equation}
    \theta = (2n + 1) \frac{\pi}{2} \hspace{0.3cm} \hbox{with} \hspace{0.3cm} n \in \boldsymbol{Z}, \hspace{0.3cm} \phi_\theta = \psi.
\end{equation}
In this case, since $\cos{(\theta)} = 0$ and $\sin{(\theta)} = 1$, we get the expression
\begin{equation}
    \begin{gathered}
        H'' = 2 \sqrt{\omega^2 - \lambda^2} K_0 + \frac{2 \omega \lambda}{\sqrt{\omega^2 - \lambda^2}} J_0 - \frac{\lambda^2}{\sqrt{\omega^2 - \lambda^2}} (K_{+}^{(b)} + K_{-}^{(b)}) - \frac{\lambda^2}{\sqrt{\omega^2 - \lambda^2}} (K_{+}^{(a)} e^{-2i \psi} + K_{-}^{(a)} e^{2i \psi}). \label{weak}
    \end{gathered}
\end{equation}
Here it is important to note that to diagonalize $H''$, it is necessary to apply a new tilting transformation in terms of the $SU(1,1)$ displacement operator of each boson $a$ and $b$, which it is defined as
\begin{equation}
    D(\xi)_{a,b} = \exp(\xi_a K^{(a)}_{+} - \xi^*_a K^{(a)}_{-}) \exp(\xi_b K^{(b)}_{+} - \xi^*_b K^{(b)}_{-}) = D(\xi_a) D(\xi_b),
\end{equation}
where $\xi_a = - \frac{\theta_a}{2} e^{-i \phi_a}$ and $\xi_b = - \frac{\theta_b}{2} e^{-i \phi_b}$. Therefore, with $H'''=D^{\dagger}(\xi)_{a,b}H''D(\xi)_{a,b}$ and $\Psi'''=D^{\dagger}(\xi)_{a,b}\Psi''$, and using the results of Eqs. (\ref{st6}) of Appendix B we obtain
\begin{equation*}
    \begin{gathered}
        H''' = 2 \sqrt{\omega^2 - \lambda^2} \left[ \frac{1}{2} \left( \cosh(2 \abs{\xi_a}) + \cosh(2 \abs{\xi_b}) \right) K_0 + \frac{1}{2} \left( \cosh(2 \abs{\xi_a}) - \cosh(2 \abs{\xi_b}) \right) J_0 \right. \\
        \left. + \frac{\sinh(2 \abs{\xi_a})}{2 \abs{\xi_a}} \left( \xi^{*}_{a} K^{(a)}_{-} + \xi_a K^{(a)}_{+} \right) + \frac{\sinh(2 \abs{\xi_b})}{2 \abs{\xi_b}} \left( \xi^{*}_{b} K^{(b)}_{-} + \xi_b K^{(b)}_{+} \right) \right] \\
        + \frac{2 \omega \lambda}{\sqrt{\omega^2 - \lambda^2}} \left[ \frac{\sinh(2 \abs{\xi_a})}{2\abs{\xi_a}} \left( \xi^*_a K^{(a)}_{-} + \xi_a K^{(a)}_{+} \right) - \frac{\sinh(2\abs{\xi_b})}{2\abs{\xi_b}} \left( \xi^*_b K^{(b)}_{-} + \xi_b K^{(b)}_{+} \right) \right. \\
        \left. + \frac{1}{2} \left( \cosh(2\abs{\xi_a}) - \cosh(2\abs{\xi_b}) \right) K_0 + \frac{1}{2} \left( \cosh(2\abs{\xi_a}) + \cosh(2\abs{\xi_b}) \right) J_0 \right]
    \end{gathered}
\end{equation*}
\begin{equation}
    \begin{gathered}
        - \frac{\lambda^2}{\sqrt{\omega^2 - \lambda^2}} \left[ \frac{\xi_b}{\abs{\xi_b}} \alpha_b K^{(b)}_0 + \beta_b \left( K_{-}^{(b)} + \frac{\xi_b}{\xi^{*}_{b}} K_{+}^{(b)} \right) + K^{(b)}_{-} \right] \\
        - \frac{\lambda^2}{\sqrt{\omega^2 - \lambda^2}} \left[ \frac{\xi^{*}_{b}}{\abs{\xi_b}} \alpha_b K^{(b)}_0 + \beta_b \left( K_{+}^{(b)} + \frac{\xi^{*}_{b}}{\xi_{b}} K_{-}^{(b)} \right) + K^{(b)}_{+} \right] \\
        - \frac{\lambda^2}{\sqrt{\omega^2 - \lambda^2}} e^{-2i \psi} \left[ \frac{\xi^{*}_{a}}{\abs{\xi_a}} \alpha_a K^{(a)}_0 + \beta_a \left( K_{+}^{(a)} + \frac{\xi^{*}_{a}}{\xi_{a}} K_{-}^{(a)} \right) + K^{(a)}_{+} \right] \\
        - \frac{\lambda^2}{\sqrt{\omega^2 - \lambda^2}} e^{2i \psi} \left[ \frac{\xi_{a}}{\abs{\xi_a}} \alpha_a K^{(a)}_0 + \beta_a \left( K_{-}^{(a)} + \frac{\xi_{a}}{\xi^{*}_{a}} K_{+}^{(a)} \right) + K^{(a)}_{-} \right].
    \end{gathered}
\end{equation}
In this expression $\alpha_a = \sinh(2 \abs{\xi_a})$, $\beta_a = \frac{1}{2} [\cosh(2 \abs{\xi_a}) - 1]$, $\alpha_b = \sinh(2 \abs{\xi_b})$ and $\beta_b = \frac{1}{2} [\cosh(2 \abs{\xi_b}) - 1]$. The coefficients of the operators $K^{(a)}_{\pm}$ and $K^{(b)}_{\pm}$ will vanish if we set the values of the coherent state parameters given by
\begin{equation}
    \begin{gathered}
        \theta_a = \tanh^{-1}{\left( \frac{- \lambda^2}{\omega^2 - \lambda^2 + \omega \lambda} \right)}, \hspace{0.3cm} \phi_a = 2 \psi, \\
        \theta_b = \tanh^{-1}{\left( \frac{- \lambda^2}{\omega^2 - \lambda^2 - \omega \lambda} \right)}, \hspace{0.3cm} \phi_b = 0.
    \end{gathered}
\end{equation}
Therefore, we obtain that the Hamiltonian $H'''$ is diagonal and is reduced to
\begin{eqnarray}
        H''' &=& \left[ \sqrt{\omega^2 - \lambda^2} [\cosh(\theta_a) + \cosh(\theta_b)] + \frac{\omega \lambda}{\sqrt{\omega^2 - \lambda^2}} [\cosh(\theta_a) - \cosh(\theta_b)] \right] K_0 \nonumber\\
        &&+ \left[ \sqrt{\omega^2 - \lambda^2} [\cosh(\theta_a) - \cosh(\theta_b)] + \frac{\omega \lambda}{\sqrt{\omega^2 - \lambda^2}} [\cosh(\theta_a) + \cosh(\theta_b)] \right] J_0 \nonumber\\
        &&+ \frac{2 \lambda^2}{\sqrt{\omega^2 - \lambda^2}} \sinh(\theta_a) K_{0}^{(a)} + \frac{2 \lambda^2}{\sqrt{\omega^2 - \lambda^2}} \sinh(\theta_b) K_{0}^{(b)}. \label{three}
\end{eqnarray}
We can write the Hamiltonian in Eq. (\ref{three}) only in terms of the operators $K_0$ and $J_0$ considering that $K_0 = K_0^{(a)} + K_0^{(b)}$ and $J_0 = K_0^{(a)} - K_0^{(b)}$. Therefore, we obtain the following result for the Hamiltonian $H'''$
\begin{equation}\label{H3}
    \begin{gathered}
        H''' = \left[ \sqrt{\omega^2 - \lambda^2} [\cosh(\theta_a) + \cosh(\theta_b)] + \frac{\omega \lambda}{\sqrt{\omega^2 - \lambda^2}} [\cosh(\theta_a) - \cosh(\theta_b)] + \frac{\lambda^2}{\sqrt{\omega^2 - \lambda^2}} [\sinh(\theta_a) + \sinh(\theta_b)] \right] K_0 \\
        + \left[ \sqrt{\omega^2 - \lambda^2} [\cosh(\theta_a) - \cosh(\theta_b)] + \frac{\omega \lambda}{\sqrt{\omega^2 - \lambda^2}} [\cosh(\theta_a) + \cosh(\theta_b)] + \frac{\lambda^2}{\sqrt{\omega^2 - \lambda^2}} [\sinh(\theta_a) - \sinh(\theta_b)] \right] J_0,
    \end{gathered}
\end{equation}
where $\cosh(\theta_a)$, $\sinh(\theta_a)$, $\cosh(\theta_b)$ and $\sinh(\theta_b)$ take the values
\begin{equation}
    \begin{gathered}
        \cosh(\theta_a) = \frac{\omega^2 - \lambda^2 + \omega \lambda}{\sqrt{(\omega^2 - \lambda^2 + \omega \lambda)^2 - \lambda^4}}, \quad\quad \sinh(\theta_a) = - \frac{\lambda^2}{\sqrt{(\omega^2 - \lambda^2 + \omega \lambda)^2 - \lambda^4}}, \\
        \cosh(\theta_b) = \frac{\omega^2 - \lambda^2 - \omega \lambda}{\sqrt{(\omega^2 - \lambda^2 - \omega \lambda)^2 - \lambda^4}}, \quad\quad \sinh(\theta_a) = - \frac{\lambda^2}{\sqrt{(\omega^2 - \lambda^2 - \omega \lambda)^2 - \lambda^4}}.
    \end{gathered}
\end{equation}
As we know, the operator $K_0$ represents the Hamiltonian of the two-dimensional harmonic oscillator, which commutes with $J_0$, so the eigenfunctions of $H'''$ turn out to be
\begin{equation}
    \Psi_{N, m} (r, \phi) = \frac{1}{\sqrt{\pi}} e^{im \phi} (-1)^{\frac{N - m}{2}} \sqrt{\frac{2 \left( \frac{N - m}{2} \right)!}{\left( \frac{N + m}{2} \right)!}} r^m L_{\frac{1}{2}(N - m)}^m (r^2) e^{- \frac{1}{2} r^2}, \label{sol1}
\end{equation}
or
\begin{equation}
    \Psi_{n, m} (r, \phi) = \frac{1}{\sqrt{\pi}} e^{im \phi} (-1)^{n_r} \sqrt{\frac{2 \left( n_r \right)!}{\left( n_r + m \right)!}} r^m L_{n}^m (r^2) e^{- \frac{1}{2} r^2}, \label{sol2}
\end{equation}
where $n = \frac{1}{2}(N - m)$ is the radial quantum number. Therefore, the eigenfunctions of the Hamiltonian of two coupled harmonic oscillators with identical free frequencies are given by
\begin{equation}
\Psi=D(\xi)D(\chi)D(\xi)_{a,b}\Psi'''=D(\xi)D(\chi)D(\xi)_{a,b} \Psi_{N, m} (r, \phi).
\end{equation}
It can be shown from the theory of irreducible representations of the $SU(1,1)$ and $SU(2)$ groups (see Appendix A) that the action of the operators $K_0$ and $J_0$ on the basis $|N,m\rangle$ are explicitly \cite{Nos1,Nos2}
\begin{equation}
K_0 \ket{N, m} = \frac{1}{2} (\hat{a}^{\dag} \hat{a} + \hat{b}^{\dag} \hat{b} + 1)\ket{N, m}=\frac{1}{2} (N + 1) \ket{N, m},\label{K0m}
\end{equation}
\begin{equation}
J_0|N,m\rangle=\frac{1}{2}\left(a^{\dag}a-b^{\dag}b\right)|N,m\rangle=\frac{m}{2}|N,m\rangle.\label{j0}
\end{equation}

Therefore, by substituting the Eqs. (\ref{K0m}) and (\ref{j0}) into Eq. (\ref{H3}) we obtain that the energy eigenvalues in the $\ket{N, m}$ basis are
\begin{equation}\label{sp}
    \begin{gathered}
        E = \left[ \sqrt{\omega^2 - \lambda^2} [\cosh(\theta_a) + \cosh(\theta_b)] + \frac{\omega \lambda}{\sqrt{\omega^2 - \lambda^2}} [\cosh(\theta_a) - \cosh(\theta_b)] + \frac{\lambda^2}{\sqrt{\omega^2 - \lambda^2}} [\sinh(\theta_a) + \sinh(\theta_b)] \right] \frac{(N + 1)}{2} \\
        + \left[ \sqrt{\omega^2 - \lambda^2} [\cosh(\theta_a) - \cosh(\theta_b)] + \frac{\omega \lambda}{\sqrt{\omega^2 - \lambda^2}} [\cosh(\theta_a) + \cosh(\theta_b)] + \frac{\lambda^2}{\sqrt{\omega^2 - \lambda^2}} [\sinh(\theta_a) - \sinh(\theta_b)] \right] \frac{m}{2}.
    \end{gathered}
\end{equation}
Here is important to note that, if we set $\lambda = 0$, we obtain that $\cosh(\theta_a) = \cosh(\theta_b) = 1$ and $\sinh(\theta_a) = \sinh(\theta_b) = 0$. With this consideration we obtain that the energy spectrum is reduced to
\begin{equation}
    E = \omega (N + 1) = \omega (2n + m + 1) = 2 \omega \left( n + \frac{m}{2} + \frac{1}{2} \right), \label{energy}
\end{equation}
which is the same energy spectrum of the two-dimensional harmonic oscillator in polar coordinates.

\section{Hamiltonian of two coupled harmonic oscillators: The anisotropic case}

In this Section we will consider the following Hamiltonian which describes the behavior of a system of two coupled harmonic oscillators with different free frequencies
\begin{equation}\label{Hani}
    H = \hbar \omega_1 \left( a^{\dag}a + \frac{1}{2} \right) + \hbar \omega_2 \left( b^{\dag}b + \frac{1}{2} \right) + \lambda e^{-i \psi} (a^{\dagger} b^{\dagger} + a^{\dagger} b) + \lambda e^{i \psi} (b^{\dagger} a + b a).
\end{equation}
Therefore, we can write this Hamiltonian (with $\hbar = 1$) as
\begin{equation}
    H = (\omega_1 + \omega_2) K_0 + (\omega_1 - \omega_2) J_0 + \lambda e^{-i \psi} (K_+ + J_+) + \lambda e^{i \psi} (K_- + J_-).
\end{equation}
From now on we will follow a procedure similar to that used in Section 2. Hence, after the first tilting transformation for this Hamiltonian in terms of the $SU(1,1)$ displacement operator $D(\xi)$ we obtain
\begin{eqnarray}
        H' &=& (\omega_1 + \omega_2) \left[ (2 \beta_\xi + 1) K_0 + \frac{\alpha_\xi \xi}{2 \abs{\xi}} K_+ + \frac{\alpha_\xi \xi^{*}}{2 \abs{\xi}} K_- \right] + (\omega_1 - \omega_2) J_0 \nonumber\\
        &&+ \lambda e^{-i \psi} \left[ \left[ \frac{\xi^*}{\abs{\xi}} \alpha_\xi K_0 + \beta_\xi \left( K_+ + \frac{\xi^*}{\xi} K_- \right) + K_+ \right] + \left[ \frac{\xi^*}{\abs{\xi}} \alpha_\xi K_{-}^{(b)} + \frac{\xi}{\abs{\xi}} \alpha_\xi K_{+}^{(a)} + (2 \beta_\xi + 1) J_+ \right] \right] \nonumber\\
        &&+ \lambda e^{i \psi} \left[ \left[ \frac{\xi}{\abs{\xi}} \alpha_\xi K_0 + \beta_\xi \left( K_- + \frac{\xi}{\xi^*} K_+ \right) + K_- \right] + \left[ \frac{\xi^*}{\abs{\xi}} \alpha_\xi K_{-}^{(a)} + \frac{\xi}{\abs{\xi}} \alpha_\xi K_{+}^{(b)} + (2 \beta_\xi + 1) J_- \right] \right].
\end{eqnarray}
In this case, the coefficients of the operators $K_{\pm}$ will vanish if we consider that the values of the coherent state parameters are
\begin{equation}
    \tau = \tanh^{-1}{\left( \frac{2 \lambda}{\omega_1 + \omega_2} \right)}, \hspace{0.3cm} \phi_\xi = \psi.
\end{equation}
Using that $\cosh{(\tau)} = \frac{\omega_1 + \omega_2}{\sqrt{(\omega_1 + \omega_2)^2 - 4 \lambda^2}}$ and $\sinh{(\tau)} = \frac{2 \lambda}{\sqrt{(\omega_1 + \omega_2)^2 - 4 \lambda^2}}$ we get the expression
\begin{eqnarray}
        H' &=& \sqrt{(\omega_1 + \omega_2)^2 - 4 \lambda^2} K_0 + (\omega_1 - \omega_2) J_0 + \frac{(\omega_1 + \omega_2) \lambda}{\sqrt{(\omega_1 + \omega_2)^2 - 4 \lambda^2}} (J_+ e^{-i \psi} + J_- e^{i \psi}) \nonumber\\
        &&- \frac{2 \lambda^2}{\sqrt{(\omega_1 + \omega_2)^2 - 4 \lambda^2}} (K_{+}^{(b)} + K_{-}^{(b)}) - \frac{2 \lambda^2}{\sqrt{(\omega_1 + \omega_2)^2 - 4 \lambda^2}} (K_{+}^{(a)} e^{-2i \psi} + K_{-}^{(a)} e^{2i \psi}).
\end{eqnarray}

Now, we can apply the tilting transformation to this Hamiltonian in terms of the $SU(2)$ displacement operator to obtain
\begin{equation}
    \begin{gathered}
        H'' = \sqrt{(\omega_1 + \omega_2)^2 - 4 \lambda^2} K_0 + (\omega_1 - \omega_2) \left[ \cos(2 \abs{\chi}) J_0 + \frac{\chi^*}{2 \abs{\chi}} \sin(2 \abs{\chi}) J_- + \frac{\chi}{2 \abs{\chi}} \sin(2 \abs{\chi}) J_+ \right] \\
        + \frac{(\omega_1 + \omega_2) \lambda}{\sqrt{(\omega_1 + \omega_2)^2 - 4 \lambda^2}} \left[ e^{-i \psi} \left[ - \frac{\chi^*}{\abs{\chi}} \alpha_\chi J_0 + \beta_\chi \left( J_+ + \frac{\chi^*}{\chi} J_- \right) + J_+ \right] + e^{i \psi} \left[ - \frac{\chi}{\abs{\chi}} \alpha_\chi J_0 + \beta_\chi \left( J_- + \frac{\chi}{\chi^*} J_+ \right) + J_- \right] \right] \\
        - \frac{2 \lambda^2}{\sqrt{(\omega_1 + \omega_2)^2 - 4 \lambda^2}} \left[ \left[ \frac{K_{-}^{(b)}}{2} \left( \cos{(2 \abs{\chi})} + 1 \right) - \frac{\chi^*}{2 \abs{\chi}} \sin{(2 \abs{\chi})} K_- - \frac{\chi^*}{2 \chi} \left( \cos{(2 \abs{\chi})} - 1 \right) K_{-}^{(a)} \right] \right. \\
        \left. + \left[ \frac{K_{+}^{(b)}}{2} \left( \cos{(2 \abs{\chi})} + 1 \right) - \frac{\chi}{2 \abs{\chi}} \sin{(2 \abs{\chi})} K_+ - \frac{\chi}{2 \chi^*} \left( \cos{(2 \abs{\chi})} - 1 \right) K_{+}^{(a)}  \right] \right] \\
        - \frac{2 \lambda^2}{\sqrt{(\omega_1 + \omega_2)^2 - 4 \lambda^2}} \left[ e^{-2i \psi} \left[ \frac{K_{+}^{(a)}}{2} \left( \cos{(2 \abs{\chi})} + 1 \right) + \frac{\chi^*}{2 \abs{\chi}} \sin{(2 \abs{\chi})} K_+ - \frac{\chi^*}{2 \chi} \left( \cos{(2 \abs{\chi})} - 1 \right) K_{+}^{(b)} \right] \right. \\
        \left. + e^{2i \psi}) \left[ \frac{K_{-}^{(a)}}{2} \left( \cos{(2 \abs{\chi})} + 1 \right) + \frac{\chi}{2 \abs{\chi}} \sin{(2 \abs{\chi})} K_- - \frac{\chi}{2 \chi^*} \left( \cos{(2 \abs{\chi})} - 1 \right) K_{-}^{(b)} \right] \right].
    \end{gathered}
\end{equation}
The coefficients of the operators $J_{\pm}$ will vanish if we set
\begin{equation}
    \theta = \tan^{-1} \left( \frac{2 \lambda (\omega_1 + \omega_2)}{(\omega_1 - \omega_2) \sqrt{(\omega_1 + \omega_2)^2 - 4 \lambda^2}} \right), \hspace{0.3cm} \phi_\theta = \psi.
\end{equation}
From these results we obtain that
\begin{equation}
    \begin{gathered}
        \cos(\theta) = \frac{(\omega_1 - \omega_2) \sqrt{(\omega_1 + \omega_2)^2 - 4 \lambda^2}}{\sqrt{4 \lambda^2 (\omega_1 + \omega_2)^2 + (\omega_1 - \omega_2)^2 [(\omega_1 + \omega_2)^2 - 4 \lambda^2]}}, \\
        \sin(\theta) = \frac{2 \lambda (\omega_1 + \omega_2)}{\sqrt{4 \lambda^2 (\omega_1 + \omega_2)^2 + (\omega_1 - \omega_2)^2 [(\omega_1 + \omega_2)^2 - 4 \lambda^2]}}.
    \end{gathered}
\end{equation}
Therefore, the Hamiltonian $H''$ can be written as
\begin{eqnarray}\nonumber
        H'' &=& \sqrt{(\omega_1 + \omega_2)^2 - 4 \lambda^2} K_0 + \sqrt{\frac{4 \lambda^2 (\omega_1 + \omega_2)^2 + (\omega_1 - \omega_2)^2 [(\omega_1 + \omega_2)^2 - 4 \lambda^2]}{(\omega_1 + \omega_2)^2 - 4 \lambda^2}} J_0 \\
        &&- \frac{2 \lambda^2}{\sqrt{(\omega_1 + \omega_2)^2 - 4 \lambda^2}} (K_{+}^{(b)} + K_{-}^{(b)}) - \frac{2 \lambda^2}{\sqrt{(\omega_1 + \omega_2)^2 - 4 \lambda^2}} (K_{+}^{(a)} e^{-2i \psi} + K_{-}^{(a)} e^{2i \psi}).
\end{eqnarray}
Applying the third tilting transformation in terms of the $SU(1,1)$ displacement operator $D(\xi)_{a,b}$ we obtain that the Hamiltonian $H'''$ is given by
\begin{equation*}
    \begin{gathered}
        H''' = \sqrt{(\omega_1 + \omega_2)^2 - 4 \lambda^2} \left[ \frac{1}{2} \left( \cosh(2 \abs{\xi_a}) + \cosh(2 \abs{\xi_b}) \right) K_0 + \frac{1}{2} \left( \cosh(2 \abs{\xi_a}) - \cosh(2 \abs{\xi_b}) \right) J_0 \right. \\
        \left. + \frac{\sinh(2 \abs{\xi_a})}{2 \abs{\xi_a}} \left( \xi^{*}_{a} K^{(a)}_{-} + \xi_a K^{(a)}_{+} \right) + \frac{\sinh(2 \abs{\xi_b})}{2 \abs{\xi_b}} \left( \xi^{*}_{b} K^{(b)}_{-} + \xi_b K^{(b)}_{+} \right) \right] \\
        + \sqrt{\frac{4 \lambda^2 (\omega_1 + \omega_2)^2 + (\omega_1 - \omega_2)^2 [(\omega_1 + \omega_2)^2 - 4 \lambda^2]}{(\omega_1 + \omega_2)^2 - 4 \lambda^2}} \left[ \frac{\sinh(2 \abs{\xi_a})}{2\abs{\xi_a}} \left( \xi^*_a K^{(a)}_{-} + \xi_a K^{(a)}_{+} \right) - \frac{\sinh(2\abs{\xi_b})}{2\abs{\xi_b}} \left( \xi^*_b K^{(b)}_{-} + \xi_b K^{(b)}_{+} \right) \right. \\
        \left. + \frac{1}{2} \left( \cosh(2\abs{\xi_a}) - \cosh(2\abs{\xi_b}) \right) K_0 + \frac{1}{2} \left( \cosh(2\abs{\xi_a}) + \cosh(2\abs{\xi_b}) \right) J_0 \right]
    \end{gathered}
\end{equation*}
\begin{equation}
    \begin{gathered}
        - \frac{2 \lambda^2}{\sqrt{(\omega_1 + \omega_2)^2 - 4 \lambda^2}} \left[ \frac{\xi_b}{\abs{\xi_b}} \alpha_b K^{(b)}_0 + \beta_b \left( K_{-}^{(b)} + \frac{\xi_b}{\xi^{*}_{b}} K_{+}^{(b)} \right) + K^{(b)}_{-} \right] \\
        - \frac{2 \lambda^2}{\sqrt{(\omega_1 + \omega_2)^2 - 4 \lambda^2}} \left[ \frac{\xi^{*}_{b}}{\abs{\xi_b}} \alpha_b K^{(b)}_0 + \beta_b \left( K_{+}^{(b)} + \frac{\xi^{*}_{b}}{\xi_{b}} K_{-}^{(b)} \right) + K^{(b)}_{+} \right] \\
        - \frac{2 \lambda^2}{\sqrt{(\omega_1 + \omega_2)^2 - 4 \lambda^2}} e^{-2i \psi} \left[ \frac{\xi^{*}_{a}}{\abs{\xi_a}} \alpha_a K^{(a)}_0 + \beta_a \left( K_{+}^{(a)} + \frac{\xi^{*}_{a}}{\xi_{a}} K_{-}^{(a)} \right) + K^{(a)}_{+} \right] \\
        - \frac{2 \lambda^2}{\sqrt{(\omega_1 + \omega_2)^2 - 4 \lambda^2}} e^{2i \psi} \left[ \frac{\xi_{a}}{\abs{\xi_a}} \alpha_a K^{(a)}_0 + \beta_a \left( K_{-}^{(a)} + \frac{\xi_{a}}{\xi^{*}_{a}} K_{+}^{(a)} \right) + K^{(a)}_{-} \right].
    \end{gathered}
\end{equation}
Now, in this case the coefficients of the operators $K^{(a)}_{\pm}$ and $K^{(b)}_{\pm}$ will vanish if we set the coherent state parameters as
\begin{equation}
    \begin{gathered}
        \theta_a = \tanh^{-1}{\left( \frac{- 4 \lambda^2}{(\omega_1 + \omega_2)^2 - 4 \lambda^2 + \sqrt{4 \lambda^2 (\omega_1 + \omega_2)^2 + (\omega_1 - \omega_2)^2 [(\omega_1 + \omega_2)^2 - 4 \lambda^2]}} \right)}, \hspace{0.3cm} \phi_a = 2 \psi, \\
        \theta_b = \tanh^{-1}{\left( \frac{- 4 \lambda^2}{(\omega_1 + \omega_2)^2 - 4 \lambda^2 - \sqrt{4 \lambda^2 (\omega_1 + \omega_2)^2 + (\omega_1 - \omega_2)^2 [(\omega_1 + \omega_2)^2 - 4 \lambda^2]}} \right)}, \hspace{0.3cm} \phi_b = 0.
    \end{gathered}
\end{equation}
Therefore, after the three tilting transformations the Hamiltonian $H'''$ is reduced to
\begin{equation}
    \begin{gathered}
        H''' = \frac{1}{2} \left[ \sqrt{(\omega_1 + \omega_2)^2 - 4 \lambda^2} [\cosh(\theta_a) + \cosh(\theta_b)] + \sqrt{\frac{4 \lambda^2 (\omega_1 + \omega_2)^2 + (\omega_1 - \omega_2)^2 [(\omega_1 + \omega_2)^2 - 4 \lambda^2]}{(\omega_1 + \omega_2)^2 - 4 \lambda^2}} [\cosh(\theta_a) - \cosh(\theta_b)] \right] K_0 \\
        + \frac{1}{2} \left[ \sqrt{(\omega_1 + \omega_2)^2 - 4 \lambda^2} [\cosh(\theta_a) - \cosh(\theta_b)] + \sqrt{\frac{4 \lambda^2 (\omega_1 + \omega_2)^2 + (\omega_1 - \omega_2)^2 [(\omega_1 + \omega_2)^2 - 4 \lambda^2]}{(\omega_1 + \omega_2)^2 - 4 \lambda^2}} [\cosh(\theta_a) + \cosh(\theta_b)] \right] J_0 \\
        + \frac{4 \lambda^2}{\sqrt{(\omega_1 + \omega_2)^2 - 4 \lambda^2}} \sinh(\theta_a) K_{0}^{(a)} + \frac{4 \lambda^2}{\sqrt{(\omega_1 + \omega_2)^2 - 4 \lambda^2}} \sinh(\theta_b) K_{0}^{(b)}. \label{tt}
    \end{gathered}
\end{equation}
The Hamiltonian of Eq. (\ref{tt}) can be expressed as a linear combination of the operators $K_0^{(a)}$ and $K_0^{(b)}$ if we consider that $K_0 = K_0^{(a)} + K_0^{(b)}$ and $J_0 = K_0^{(a)} - K_0^{(b)}$. Thus,
\begin{equation}\label{H3ani}
    \begin{gathered}
        H''' = \frac{1}{2} \left[ \sqrt{(\omega_1 + \omega_2)^2 - 4 \lambda^2} [\cosh(\theta_a) + \cosh(\theta_b)] + \sqrt{\frac{4 \lambda^2 (\omega_1 + \omega_2)^2 + (\omega_1 - \omega_2)^2 [(\omega_1 + \omega_2)^2 - 4 \lambda^2]}{(\omega_1 + \omega_2)^2 - 4 \lambda^2}} [\cosh(\theta_a) - \cosh(\theta_b)] \right. \\
        \left. + \frac{4 \lambda^2}{\sqrt{(\omega_1 + \omega_2)^2 - 4 \lambda^2}} [\sinh(\theta_a) + \sinh(\theta_b)] \right] K_0 \\
        + \frac{1}{2} \left[ \sqrt{(\omega_1 + \omega_2)^2 - 4 \lambda^2} [\cosh(\theta_a) - \cosh(\theta_b)] + \sqrt{\frac{4 \lambda^2 (\omega_1 + \omega_2)^2 + (\omega_1 - \omega_2)^2 [(\omega_1 + \omega_2)^2 - 4 \lambda^2]}{(\omega_1 + \omega_2)^2 - 4 \lambda^2}} [\cosh(\theta_a) + \cosh(\theta_b)] \right. \\
        \left. + \frac{4 \lambda^2}{\sqrt{(\omega_1 + \omega_2)^2 - 4 \lambda^2}} [\sinh(\theta_a) - \sinh(\theta_b)] \right] J_0,
    \end{gathered}
\end{equation}
where $\cosh(\theta_a)$, $\sinh(\theta_a)$, $\cosh(\theta_b)$ and $\sinh(\theta_b)$ take the following expressions
\begin{equation}
    \begin{gathered}
        \cosh(\theta_a) = \frac{(\omega_1 + \omega_2)^2 - 4 \lambda^2 + \sqrt{4 \lambda^2 (\omega_1 + \omega_2)^2 + (\omega_1 - \omega_2)^2 [(\omega_1 + \omega_2)^2 - 4 \lambda^2]}}{\sqrt{\left[ (\omega_1 + \omega_2)^2 - 4 \lambda^2 + \sqrt{4 \lambda^2 (\omega_1 + \omega_2)^2 + (\omega_1 - \omega_2)^2 [(\omega_1 + \omega_2)^2 - 4 \lambda^2]} \right]^2 - 16 \lambda^4}}, \\
        \sinh(\theta_a) = - \frac{4 \lambda^2}{\sqrt{\left[ (\omega_1 + \omega_2)^2 - 4 \lambda^2 + \sqrt{4 \lambda^2 (\omega_1 + \omega_2)^2 + (\omega_1 - \omega_2)^2 [(\omega_1 + \omega_2)^2 - 4 \lambda^2]} \right]^2 - 16 \lambda^4}}, \\
        \cosh(\theta_b) = \frac{(\omega_1 + \omega_2)^2 - 4 \lambda^2 - \sqrt{4 \lambda^2 (\omega_1 + \omega_2)^2 + (\omega_1 - \omega_2)^2 [(\omega_1 + \omega_2)^2 - 4 \lambda^2]}}{\sqrt{\left[ (\omega_1 + \omega_2)^2 - 4 \lambda^2 - \sqrt{4 \lambda^2 (\omega_1 + \omega_2)^2 + (\omega_1 - \omega_2)^2 [(\omega_1 + \omega_2)^2 - 4 \lambda^2]} \right]^2 - 16 \lambda^4}}, \\
        \sinh(\theta_a) = - \frac{4 \lambda^2}{\sqrt{\left[ (\omega_1 + \omega_2)^2 - 4 \lambda^2 - \sqrt{4 \lambda^2 (\omega_1 + \omega_2)^2 + (\omega_1 - \omega_2)^2 [(\omega_1 + \omega_2)^2 - 4 \lambda^2]} \right]^2 - 16 \lambda^4}}.
    \end{gathered}
\end{equation}
Therefore, just as in the isotropic case, the eigenfunctions for the Hamiltonian of Eq. (\ref{Hani}) are given by
\begin{equation}
\Psi=D(\xi)D(\chi)D(\xi)_{a,b}\Psi'''=D(\xi)D(\chi)D(\xi)_{a,b} \Psi_{N, m} (r, \phi),
\end{equation}
and the energy spectrum in the $\ket{N,m}$ basis is obtained by substituting the results of Eqs. (\ref{K0m}) and (\ref{j0}) into Eq. (\ref{H3ani}) to obtain
\begin{equation}\label{spani}
    \begin{gathered}
        E = \frac{1}{2} \left[ \sqrt{(\omega_1 + \omega_2)^2 - 4 \lambda^2} [\cosh(\theta_a) + \cosh(\theta_b)] + \sqrt{\frac{4 \lambda^2 (\omega_1 + \omega_2)^2 + (\omega_1 - \omega_2)^2 [(\omega_1 + \omega_2)^2 - 4 \lambda^2]}{(\omega_1 + \omega_2)^2 - 4 \lambda^2}} [\cosh(\theta_a) - \cosh(\theta_b)] \right. \\
        \left. + \frac{4 \lambda^2}{\sqrt{(\omega_1 + \omega_2)^2 - 4 \lambda^2}} [\sinh(\theta_a) + \sinh(\theta_b)] \right] \frac{(N + 1)}{2} \\
        + \frac{1}{2} \left[ \sqrt{(\omega_1 + \omega_2)^2 - 4 \lambda^2} [\cosh(\theta_a) - \cosh(\theta_b)] + \sqrt{\frac{4 \lambda^2 (\omega_1 + \omega_2)^2 + (\omega_1 - \omega_2)^2 [(\omega_1 + \omega_2)^2 - 4 \lambda^2]}{(\omega_1 + \omega_2)^2 - 4 \lambda^2}} [\cosh(\theta_a) + \cosh(\theta_b)] \right. \\
        \left. + \frac{4 \lambda^2}{\sqrt{(\omega_1 + \omega_2)^2 - 4 \lambda^2}} [\sinh(\theta_a) - \sinh(\theta_b)] \right] \frac{m}{2}.
    \end{gathered}
\end{equation}
Finally, it is easy to show that if we set $\omega_1 = \omega_2$, the energy spectrum of Eq. (\ref{spani}) reduces to that obtained in Eq. (\ref{sp}) of Section 2 for the isotropic case. Moreover, if we consider $\lambda = 0$ and $\omega_1 = \omega_2$, we obtain that $\cosh(\theta_a) = \cosh(\theta_b) = 1$ and $\sinh(\theta_a) = \sinh(\theta_b) = 0$. With these considerations the energy spectrum of Eq. (\ref{spani}) is reduced to that of the
two dimensional harmonic oscillator (Eq. (\ref{energy})), as it was expected.

\section{Concluding remarks}

In this work we used a purely algebraic approach based on the theory of the groups $SU(1,1)$ and $SU(2)$ to study the Hamiltonian of two coupled harmonic oscillators. Within this problem, both the isotropic and the anisotropic cases were studied.

To obtain the energy spectrum and the eigenfunctions of each case, the Hamiltonian was diagonalized by applying three different similarity transformations based on the displacement operator of the $SU(1,1)$ and $SU(2)$ groups. By making some considerations to the transformation parameters, we were able to transform the original Hamiltonian to one which depends only on the Hamiltonian of the two-dimensional harmonic oscillator and the number difference operator $J_0$. This allowed us to calculate the energy spectrum and eigenfunctions for both the isotropic and anisotropic cases.

Therefore, a general Hamiltonian of two coupled oscillators which is initially not easy to study, was solved exactly by an elegant, purely algebraic diagonalization process. Thus, the original Hamiltonians for the isotropic and anisotropic cases were transformed into the most important elementary problem in physics, the harmonic oscillator. Furthermore, the tilting transformation method has been successfully used to diagonalize and exactly solve other problems in quantum mechanics and quantum optics, as can be seen in Refs. \cite{Nos3,Nos4,Nos5,Nos6}.

\section*{Acknowledgments}

This work was partially supported by SNII-M\'exico, EDI-IPN, SIP-IPN Project Number $20241764$.\\

\section*{Disclosures}

The authors declare no conflicts of interest.

\section*{Data Availability}

No data were generated or analyzed in the presented research.

\renewcommand{\theequation}{A.\arabic{equation}}
\setcounter{equation}{0}

\section*{Appendix A. The $SU(1,1)$ and $SU(2)$ group theory and its Perelomov coherent states}

The $su(1,1)$ and $su(2)$ Lie algebras satisfy the following commutation relations \cite{Vourdas}
\begin{eqnarray}
    [K_{0}, K_{\pm}] = \pm K_{\pm}, \quad\quad [K_{-}, K_{+}] = 2 K_{0}, \label{algebra1}
\end{eqnarray}
\begin{eqnarray}
    [J_{0}, J_{\pm}] = \pm J_{\pm}, \quad\quad [J_{+}, J_{-}] = 2 J_{0}. \label{algebra2}
\end{eqnarray}
In these expressions, the operators $K_{\pm}$, $K_0$ are the generators of the $su(1,1)$ Lie algebra, while the operators $J_\pm$ and $J_{0}$ are the generators of the $su(2)$ Lie algebra. The Casimir operators $K^2$ and $J^{2}$ for these algebras satisfy $[K^{2},K_{\pm}]=[K^{2},K_{0}]=0$ and $[J^{2},J_{\pm}]=[J^{2},J_{0}]=0$, and have the form
\begin{equation}
    K^2 = K_0^2 - \frac{1}{2} \left(K_+K_- + K_-K_+ \right), \quad\quad J^{2} = J_0^2 + \frac{1}{2} \left(J_+J_- + J_-J_+ \right).
\end{equation}
Now, the discrete representation of the $su(1,1)$ and $su(2)$ Lie algebra are given by
\begin{align}
    &K_{+} |k, n\rangle = \sqrt{(n + 1)(2k + n)} |k, n + 1 \rangle, \hspace{1.9cm} J_{+} |j, \mu \rangle = \sqrt{(j - \mu)(j + \mu + 1)} |j, \mu + 1 \rangle, \label{k+n}\\
    &K_{-} |k, n \rangle = \sqrt{n (2k + n - 1)} |k, n - 1 \rangle, \hspace{2.2cm} J_{-} |j, \mu \rangle = \sqrt{(j + \mu) (j - \mu + 1)} |j, \mu - 1 \rangle, \label{k-n}\\
    &K_{0} |k, n \rangle = (k + n) |k, n \rangle, \hspace{4.2cm} J_{0} |j, \mu \rangle = \mu |j, \mu \rangle, \label{k0n}\\
    &K^2 |k, n \rangle = k (k - 1) |k, n \rangle, \hspace{4cm} J^2 |j, \mu \rangle = j (j + 1) |j, \mu \rangle. \label{Cas}
\end{align}

The displacement operators $D(\xi)$ and $D(\chi)$ for these algebras are defined in terms of the creation and annihilation operators $\{K_+, K_- \}$ and $\{J_+,J_- \}$ as
\begin{equation}
    D(\xi)_{su(1,1)} = \exp(\xi K_{+} - \xi^{*} K_{-}), \quad\quad D(\chi)_{su(2)} = \exp(\chi J_{+} - \chi^{*} J_{-}), \label{do}
\end{equation}
where $\xi = - \rfrac{1}{2} \tau e^{-i \phi_\xi}$ and $\chi = - \rfrac{1}{2} \theta e^{-i \phi_\theta}$ with $- \infty < \tau, \theta < \infty$ and $0 \leq \phi_\xi, \phi_\chi \leq 2 \pi$. \\
Hence, the $SU(1,1)$ Perelomov number coherent states are defined as the action of the operator $D(\xi)$ onto an arbitrary state $|k, n \rangle$ as \cite{Nos1}
\begin{eqnarray}
    |\zeta_{\xi}, k, n \rangle & = & \sum_{s = 0}^\infty \frac{\zeta_{\xi}^s}{s!} \sum_{j = 0}^n \frac{(-\zeta_{\xi}^*)^j}{j!} e^{\eta_{\xi} (k + n - j)} \frac{\sqrt{\Gamma(2k + n) \Gamma(2k + n - j + s)}}{\Gamma(2k + n - j)} \nonumber\\
    &&\times \frac{\sqrt{\Gamma(n + 1) \Gamma(n - j + s + 1)}}{\Gamma(n - j + 1)} |k, n - j + s \rangle. \label{PNCS}
\end{eqnarray}
In an analogous way, the $SU(2)$ Perelomov number coherent states are defined as $D(\chi)|j ,\mu \rangle$ and explicitly are given by \cite{Nos2}
\begin{eqnarray}
    |\zeta_{\chi}, j ,\mu \rangle & = & \sum_{s = 0}^{j - \mu + n} \frac{\zeta_{\chi}^{s}}{s!} \sum_{n = 0}^{\mu + j} \frac{(-\zeta_{\chi}^*)^{n}}{n!} e^{\eta_{\chi} (\mu - n)} \frac{\Gamma(j - \mu + n + 1)}{\Gamma(j + \mu - n + 1)} \nonumber\\
    && \times \left[\frac{\Gamma(j + \mu + 1) \Gamma(j + \mu - n + s + 1)}{\Gamma(j - \mu + 1) \Gamma(j - \mu + n - s + 1)} \right]^{\frac{1}{2}} |j, \mu - n + s \rangle. \label{PNCS2}
\end{eqnarray}
where $\zeta_{\xi} = -\tanh( \frac{\tau}{2} ) e^{-i \phi_{\xi}}$, $\eta_{\xi} = \ln( 1 - |\zeta_{\xi}|^{2} )$ and $\zeta_{\chi} = -\tan( \frac{\theta}{2} ) e^{-i \phi_{\chi}}$ and $\eta_{\chi} = \ln( 1 + |\zeta_{\chi}|^{2} )$.

On the other hand, as it is well known, the bosonic annihilation $\hat{a}$, $\hat{b}$ and creation $\hat{a}^{\dag}$, $\hat{b}^{\dag}$ operators obey the commutation relations
\begin{equation}
    [\hat{a}, \hat{a}^{\dag}] = [\hat{b}, \hat{b}^{\dag}] = 1,
\end{equation}
\begin{equation}
    [\hat{a}, \hat{b}] = [\hat{a}^{\dag}, \hat{b}^{\dag}] = [\hat{a}^{\dag}, \hat{b}] = [\hat{a}, \hat{b}^{\dag}] = 0. \label{boson}
\end{equation}
These operators can be used appropriately to construct realizations of the $su(2)$ and $su(1,1)$ algebras. Thus, the $su(2)$ Lie algebra realization is given by the four operators \cite{Vourdas2}
\begin{equation}
    J_+ = \hat{a}^{\dag} \hat{b}, \quad\quad J_- = \hat{b}^{\dag} \hat{a}, \quad\quad J_0 = \frac{1}{2} (\hat{a}^{\dag} \hat{a} - \hat{b}^{\dag} \hat{b}), \quad\quad J^{2} = \frac{1}{4} N (N + 2), \label{su2}
\end{equation}
where $N = \hat{a}^{\dag} \hat{a} + \hat{b}^{\dag} \hat{b}$. \\
In terms of the $su(1,1)$ Lie algebra, we can obtain two different realizations \cite{Vourdas2}
\begin{equation}
    K_{+} = \hat{a}^{\dag} \hat{b}^{\dag}, \quad\quad K_{-} = \hat{b} \hat{a}, \quad\quad K_{0} = \frac{1}{2} (\hat{a}^{\dag} \hat{a} + \hat{b}^{\dag} \hat{b} + 1), \quad \quad K^{2} = J_{0}^{2} - \frac{1}{4}, \label{su11ab}
\end{equation}
and
\begin{equation}
    K_{+}^{(a)} = \frac{1}{2} \hat{a}^{\dag}{}^{2}, \quad\quad K_{-}^{(a)} = \frac{1}{2} \hat{a}^{2}, \quad\quad K_{0}^{(a)} = \frac{1}{2} \left( \hat{a}^{\dag} \hat{a} + \frac{1}{2} \right), \quad\quad K_{(a)}^{2} = -\frac{3}{16}. \label{su11a}
\end{equation}
Notice that for the second realization of the $su(1,1)$ Lie algebra, the Casimir operator $K_{(a)}^{2}$ is constant and the Bargmann index $k$ only can take the values $k=\frac{1}{4}$ and $k=\frac{3}{4}$.

\renewcommand{\theequation}{B.\arabic{equation}}
\setcounter{equation}{0}

\section*{Appendix B. Similarity transformations of the $SU(1,1)$ and $SU(2)$ group generators}

The $SU(1,1)$ and $SU(2)$ displacement operators defined in Appendix A and we use the Baker-Campbell-Hausdorff identity
\begin{equation}
    e^{-A} B e^{A} = B + [B, A] + \frac{1}{2!} [[B, A], A] + \frac{1}{3!} [[[B, A], A], A] + ...,
\end{equation}
can be used to compute similarity transformations if they act on the $SU(1,1)$ and $SU(2)$ group generators. Thus, by using the $SU(1,1)$ displacement operator $D(\xi) = \exp(\xi K_+ - \xi^{*} K_-)$ and the commutation relations of Eqs. (\ref{algebra1}) and (\ref{algebra2}), we can compute the following similarity transformations for the $SU(1,1)$ operators $\{K_0,K_{\pm}\}$
\begin{equation}\label{st1}
    \begin{gathered}
        D^{\dagger}(\xi) K_0 D(\xi) = (2 \beta_\xi + 1) K_0 + \frac{\alpha_\xi \xi}{2 \abs{\xi}} K_+ + \frac{\alpha_\xi \xi^{*}}{2 \abs{\xi}} K_-, \\
        D^{\dagger}(\xi) K_+ D(\xi) = \frac{\xi^*}{\abs{\xi}} \alpha_\xi K_0 + \beta_\xi \left( K_+ + \frac{\xi^*}{\xi} K_- \right) + K_+, \\
        D^{\dagger}(\xi) K_- D(\xi) = \frac{\xi}{\abs{\xi}} \alpha_\xi K_0 + \beta_\xi \left( K_- + \frac{\xi}{\xi^*} K_+ \right) + K_-.
    \end{gathered}
\end{equation}
The similarity transformations for the $SU(2)$ group generators $\{J_0,J_{\pm}\}$ in terms of the $SU(1,1)$ displacement operator $D(\xi)$ are given by
\begin{equation} \label{st2}
    \begin{gathered}
        D^{\dagger}(\xi) J_0 D(\xi) = J_0, \\
        D^{\dagger}(\xi) J_+ D(\xi) = \frac{\xi^*}{\abs{\xi}} \alpha_\xi K_{-}^{(b)} + \frac{\xi}{\abs{\xi}} \alpha_\xi K_{+}^{(a)} + (2 \beta_\xi + 1) J_+, \\
        D^{\dagger}(\xi) J_- D(\xi) = \frac{\xi^*}{\abs{\xi}} \alpha_\xi K_{-}^{(a)} + \frac{\xi}{\abs{\xi}} \alpha_\xi K_{+}^{(b)} + (2 \beta_\xi + 1) J_-,
    \end{gathered}
\end{equation}
In these expressions $\alpha_\xi = \sinh{(2\abs{\xi})}$ and $\beta_\xi = \rfrac{1}{2} [ \cosh{(2\abs{\xi})} - 1]$.

Analogously, the $SU(2)$ displacement operator $D(\chi) = \exp(\chi J_+ - \chi^{*} J_-)$ can be used to obtain similarity transformations for the $SU(1,1)$ and $SU(2)$ Schwinger realizations of Eqs. (\ref{su2}), (\ref{su11ab}) and (\ref{su11a}). Therefore, for the $SU(2)$ Schwinger realization of Eq. (\ref{su2}) we obtain
\begin{equation}\label{st3}
    \begin{gathered}
        D^{\dagger}(\chi) J_0 D(\chi) = (2 \beta_\chi + 1) J_{0} + \frac{\chi}{2 |\chi|} \alpha_\chi J_{+} + \frac{\chi^{*}}{2 |\chi|} \alpha_\chi J_{-}, \\
        D^{\dagger}(\chi) J_+ D(\chi) = - \frac{\chi^*}{\abs{\chi}} \alpha_\chi J_0 + \beta_\chi \left( J_+ + \frac{\chi^*}{\chi} J_- \right) + J_+, \\
        D^{\dagger}(\chi) J_- D(\chi) = - \frac{\chi}{\abs{\chi}} \alpha_\chi J_0 + \beta_\chi \left( J_- + \frac{\chi}{\chi^*} J_+ \right) + J_-.
    \end{gathered}
\end{equation}
For the two-boson $SU(1,1)$ Schwinger realization (\ref{su11ab}) we can compute the following $SU(2)$ similarity transformations
\begin{equation}\label{st4}
    \begin{gathered}
        D^{\dagger}(\chi) K_0 D(\chi) = K_0, \\
        D^{\dagger}(\chi) K_+ D(\chi) = (2 \beta_\chi + 1) K_+ - \frac{\chi}{\abs{\chi}} \alpha_\chi K_+^{(a)} + \frac{\chi^{*}}{\abs{\chi}} \alpha_\chi K_+^{(b)}, \\
        D^{\dagger}(\chi) K_- D(\chi) = (2 \beta_\chi + 1) K_- - \frac{\chi^{*}}{\abs{\chi}} \alpha_\chi K_-^{(a)} + \frac{\chi}{\abs{\chi}} \alpha_\chi K_-^{(b)}.
    \end{gathered}
\end{equation}
We can obtain the $SU(2)$ similarity transformations for the one-boson $SU(1,1)$ Schwinger realization (\ref{su11a}). Therefore, if we assume that we have two realizations in terms of the $a$ and $b$ bosons
\begin{equation}\label{st5}
    \begin{gathered}
        D^{\dagger}(\chi) K_+^{(a)} D(\chi) = \left( \beta_\chi + 1 \right) K_{+}^{(a)} + \frac{\chi^*}{2 \abs{\chi}} \alpha_\chi K_+ - \frac{\chi^*}{\chi} \beta_\chi K_{+}^{(b)}, \\
        D^{\dagger}(\chi) K_-^{(a)} D(\chi) = \left( \beta_\chi + 1 \right) K_{-}^{(a)} + \frac{\chi}{2 \abs{\chi}} \alpha_\chi K_- - \frac{\chi}{\chi^*} \beta_\chi K_{-}^{(b)}, \\
        D^{\dagger}(\chi) K_+^{(b)} D(\chi) = \left( \beta_\chi + 1 \right) K_{+}^{(b)} - \frac{\chi}{2 \abs{\chi}} \alpha_\chi K_+ - \frac{\chi}{\chi^*} \beta_\chi K_{+}^{(a)}, \\
        D^{\dagger}(\chi) K_-^{(b)} D(\chi) = \left( \beta_\chi + 1 \right) K_{-}^{(b)} - \frac{\chi^*}{2 \abs{\chi}} \alpha_\chi K_- - \frac{\chi^*}{\chi} \beta_\chi K_{-}^{(a)}.
    \end{gathered}
\end{equation}
Here we have $\alpha_\chi = \sin{(2\abs{\chi})}$ and $\beta_\chi = \rfrac{1}{2} [ \cos{(2\abs{\chi})} - 1]$. \\
Finally, we can define $D(\xi)_{a,b} = \exp(\xi_a K^{(a)}_{+} - \xi^*_a K^{(a)}_{-})\exp(\xi_b K^{(b)}_{+} - \xi^*_b K^{(b)}_{-}) = D(\xi_a) D(\xi_b)$, which is a general $SU(1,1)$ displacement operator of each boson $a$ and $b$, where $\xi_a = - \frac{\theta_a}{2} e^{-i \phi_a}$ and $\xi_b = - \frac{\theta_b}{2} e^{-i \phi_b}$. The similarity transformations with this operator of the operators $\{J_0,K_0,K_{\pm}^{(a)},K_{\pm}^{(b)}\}$ are explicitly given by
\begin{equation}\label{st6}
    \begin{gathered}
        D^{\dagger} (\xi)_{a,b} J_0 D(\xi)_{a,b} = \frac{1}{2} \left[ \cosh(2 \abs{\xi_a}) + \cosh(2 \abs{\xi_b}) \right] J_0 + \frac{1}{2} \left[ \cosh(2 \abs{\xi_a}) - \cosh(2 \abs{\xi_b}) \right] K_0 \\
        + \frac{\sinh(2 \abs{\xi_a})}{2 \abs{\xi_a}} \left( \xi_{a}^{*} K^{(a)}_- + \xi_a K^{(a)}_+ \right) - \frac{\sinh(2 \abs{\xi_b})}{2 \abs{\xi_b}} \left( \xi_{b}^{*} K^{(b)}_- + \xi_b K^{(b)}_+ \right), \\
        D^{\dagger} (\xi)_{a,b} K_0 D(\xi)_{a,b} = \frac{1}{2} \left[ \cosh(2 \abs{\xi_a}) + \cosh(2 \abs{\xi_b}) \right] K_0 + \frac{1}{2} \left[ \cosh(2 \abs{\xi_a}) - \cosh(2 \abs{\xi_b}) \right] J_0 \\
        + \frac{\sinh(2 \abs{\xi_a})}{2 \abs{\xi_a}} \left( \xi_{a}^{*} K^{(a)}_- + \xi_a K^{(a)}_+ \right) + \frac{\sinh(2 \abs{\xi_b})}{2 \abs{\xi_b}} \left( \xi_{b}^{*} K^{(b)}_- + \xi_b K^{(b)}_+ \right), \\
        D^{\dagger} (\xi)_{a,b} K_+^{(a)} D(\xi)_{a,b} = \frac{\xi_{a}^{*}}{\abs{\xi_a}} \alpha_a K^{(a)}_{0} + \beta_a \left( K^{(a)}_{+} + \frac{\xi_{a}^{*}}{\xi_{a}} K^{(a)}_{-} \right) + K^{(a)}_{+}, \\
        D^{\dagger} (\xi)_{a,b} K_-^{(a)} D(\xi)_{a,b} = \frac{\xi_a}{\abs{\xi_a}} \alpha_a K_0^{(a)} + \beta_a \left( K^{(a)}_- + \frac{\xi_a}{\xi_{a}^{*}} K^{(a)}_+ \right) + K^{(a)}_-, \\
        D^{\dagger} (\xi)_{a,b} K_+^{(b)} D(\xi)_{a,b} = \frac{\xi_{b}^{*}}{\abs{\xi_b}} \alpha_b K^{(b)}_{0} + \beta_b \left( K^{(b)}_{+} + \frac{\xi_{b}^{*}}{\xi_{b}} K^{(b)}_{-} \right) + K^{(b)}_{+}, \\
        D^{\dagger} (\xi)_{a,b} K_-^{(b)} D(\xi)_{a,b} = \frac{\xi_b}{\abs{\xi_b}} \alpha_b K_0^{(b)} + \beta_b \left( K^{(b)}_- + \frac{\xi_b}{\xi_{b}^{*}} K^{(b)}_+ \right) + K^{(b)}_-,
    \end{gathered}
\end{equation}
where $\alpha_a = \sinh(2 \abs{\xi_a})$, $\beta_a = \frac{1}{2} [\cosh(2 \abs{\xi_a}) - 1]$, $\alpha_b = \sinh(2 \abs{\xi_b})$ and $\beta_b = \frac{1}{2} [\cosh(2 \abs{\xi_b}) - 1]$.

\end{document}